\documentclass[prl,aps,showpacs,twocolumn]{revtex4}
\usepackage{epsfig}
\usepackage{bm}

\begin{document}

\preprint{HEP/123-qed}

\title{Transverse velocities, intermittency and asymmetry
in fully developed turbulence}
\author{\normalsize{S. I. Vainshtein}}
\affiliation{\small {\it Department of Astronomy and Astrophysics, University
of Chicago, Chicago,  60637, USA}}

\date{\today}

\begin{abstract}
\normalsize{
Using experimental transverse velocities data for very high Reynolds number
turbulence, we suggest a model describing both formation of intermittency
and asymmetry of turbulence. 
The model, called "bump-model" is a modification of ramp-model suggested
earlier \cite{94}. The connection between asymmetry and intermittency  makes it 
possible to study the latter with relatively low moments.
}
\end{abstract}

\maketitle

Self-similar properties of turbulence, suggested by Kolmogorov
\cite{k41}, have been intensively studied for a long time. The theory predicted
simple scaling for the longitudinal velocity increments $u_r =u(x+r)-u(x)$, namely,
$\langle |u_r|^p\rangle 
\sim r^{p/3}$. It became clear, however, that there
are  corrections to these scalings, usually attributed  to 
intermittency. The only exception is the so-called $4/5$th-Kolmogorov law \cite{law} which is
exact in inertial range. According to the law, the third 
moment, that is the structure function, 
$\langle u_r^3\rangle=B_{uuu}(r)=-4/5 \varepsilon r$ in inertial range, and this scaling 
has no intermittency corrections \cite{Monin}.
Nevertheless, more detailed study
of this third order structure function proved to be useful in understanding
the  intermittency. 
It is indeed important to understand what contribution into the third moment give the
tails. It is natural to assume that the main events from the core of $P(u_r)$, the
PDF,   mainly  contribute. This is certainly the case for even order structure
functions, or for moments like $\langle |u_r|^p\rangle$. As to the odd order moments, 
we note that $\langle u_r^3\rangle\not=0$, while $\langle u_r\rangle=0$. This means that
the PDF is asymmetric. However, the core of the PDF may be more or less symmetric, in 
which case the contribution of the tails would be substantial.
And indeed, this asymmetry, described by a ramp-model, also
 suggest intermittency (in addition to the asymmetry) \cite{94}, \cite{96}. 
Further studies showed that indeed the tails of $P(u_r)$, responsible for the
intermittency, give a
substantial contribution to $B_{uuu}$ \cite{asym}. Another way to check this connection
between asymmetry and intermittency is to compare directly the PDF for positive and 
negative parts of $u_r$, and we can see that the asymmetry of the PDF stretches far into
the tails \cite{asym}. 
 The transverse velocities give
additional information about both asymmetry and intermittency, and this paper is devoted to
their study. 

\noindent
The transverse (vertical) component of the velocity increments 
$v_r=v(x+r)-v(x)$ is also supposed
to possess asymmetry, although $\langle v_r^3\rangle=0$. Assuming
isotropic turbulence,  the only non-vanishing correlation is 
(see \cite{book}, \cite{Monin})
\begin{equation}
B_{uvv}=\langle u_rv_r^2 \rangle=\frac{1}{6}\frac{d(B_{uuu}r)}{dr}.
\label{connection}
\end{equation}
We used data  acquired at Brookhaven National Lab for longitudinal and transverse components
of the velocity (40 million samples of each, courtesy of Sreenivasan). The estimated Taylor 
Reynolds number is 10680. 
As seen from Fig. 1, experimental $B_{uvv}$ is close to that obtained from
(\ref{connection}), especially at small distances between the points, in 
agreement with earlier observations \cite{Susan} (see their Fig. 2).
Indeed at small scales the statistical properties are more isotropic, in accordance with
Kolmogorov ideas about local isotropy. 
\begin{figure}[h]
\psfig{file=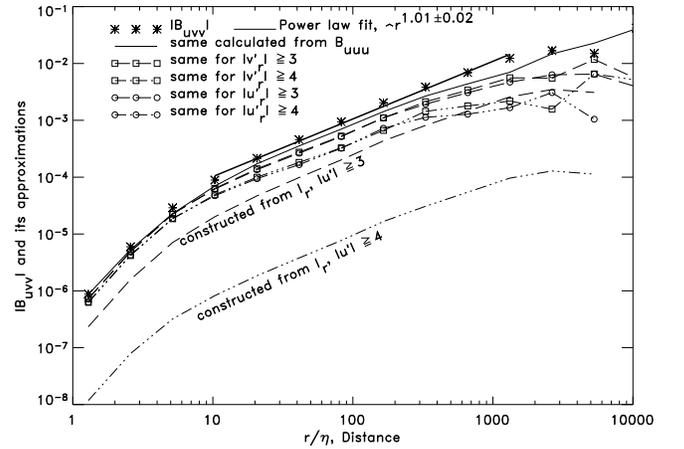,width=3.5in}
\caption
{Different third order structure functions: experimental $\langle u_rv_r^2\rangle$,
and its cumulative moments. The distance is given in terms of Kolmogorov microscale $\eta$.
}
\end{figure}

Denote $u'_r=u_r/\sigma_u$, $v'_r=v_r/\sigma_v$,  and $\sigma_u=
\langle u_r^2\rangle^{1/2}, ~\sigma_v=\langle v_r^2\rangle^{1/2}$. 
We will consider  cumulative moments, 
$$\langle u'_rv'^2_r\rangle\rule[-2mm]{.5mm}
{6mm}_{ |u'_r| \ge t}=
 \left(\int_{-\infty}^{-t} +\int_t^\infty\right)du'_r
\int_{-\infty}^\infty dv'_r
u'_rv'^2_rP(u'_r,v'_r),$$
$$\langle u'_rv'^2_r\rangle\rule[-2mm]{.5mm}{6mm}_{|v'_r| \ge t}=
\int_{-\infty}^\infty du'_r \left(\int_{-\infty}^{-t} +\int_t^\infty\right)
 dv'_r u'_rv'^2_rP(u'_r,v'_r)
$$
where $P(u'_r,v'_r)$ is the distribution function, and $t$ is a 
number. If $t\ll 1$, then essentially the whole distribution  works, and
the cumulative moments are expected almost to coincide with $\langle 
u'_rv'^2_r\rangle=k(r)$, where $k=B_{uvv}/(\sigma_u\sigma_v^2)$, - analog to
skewness. For not  
small $t$, we are dealing with the tails of the distribution, and it is important to
know what contribution they give to the distribution. Figure 1 shows these moments
for $t= 3$ or $4$. It can be seen that the moments thus constructed do not
deviate much from the experimental $B_{uvv}(r)$.

In order to have some comparison with a ``regular" behavior when the tails
are absent, we constructed a series of PDF's $I_r(u')$ for different distances $r$,
 as a sum of two Gaussian functions, and 
satisfying $\langle u'^0\rangle_I=\int I_rdu'=1$, $\langle u'\rangle_I=\int I_ru'du'=0$, 
$\langle u'^2\rangle_I=1$, and
$\langle u'^3\rangle_I=k(r)$. 
 Then, obviously, $\langle u_r\rangle_I=\langle u'\rangle_I
\sigma_u=0$, $\langle v_r\rangle_I=\langle u'\rangle_I\sigma_v=0$, $\langle u_r^2\rangle_I
=B_{uu}$, $\langle v_r^2\rangle_I
=B_{vv}$, and $\langle u_rv_r^2\rangle_I=B_{uvv}$. 
We now reconstruct the third moment for cumulative
average $\langle u'^3\rangle_I\rule[-2mm]{.5mm}{6mm}_{|u'| \ge t}$, $t=3$ or $4$.
Corresponding moments are depicted in Fig. 1. We note that even for $t=3$, the
cumulative moment constructed from $I_r$ is essentially lower than $B_{uvv}$; 
only for large distances
it mixes with the experimental cumulative moments. As to the case $t=4$, it 
can be seen that
the difference between the $I_r$-cumulative moments and experimental moments
is dramatic. We conclude that the contribution of the tails for experimental data
 is substantial, which can be seen when comparing the experimental cumulative moments with
real moments, -- on one hand, and, on the other hand, comparing them 
with those constructed from $I_r$ -- that does not contain any tails by definition.

As a non-vanishing $B_{uuu}$ is a result of asymmetry of the PDF for $u_r$, the correlation
$B_{uvv}$, obeying 
(\ref{connection}), is therefore related to the asymmetry. Indeed,
 in order that $B_{uvv}< 0$, there should be an anti-correlation between $u_r$ and
$v_r^2$, -- decreasing $u_r$ is accompanied by increasing $v_r^2$, and {\it vice versa}. Roughly
speaking, the conditional average $B_{uvv}(u_r<0)>B_{uvv}(u_r>0)$. If the asymmetry is indeed
related to intermittency, this conditional inequality should be satisfied for $u_r>t$ versus
$u_r<-t$, where $t$ is not a small number. To check this 
 we consider, first, distributions 
for smallest $r$'s corresponding to the distance between two neighbor samples. Second, we consider
cumulative moments, $\langle u_r v_r^2\rangle\rule[-2mm]{.5mm}{6mm}_{u_r\le -t}$, and 
$\langle u_r v_r^2\rangle\rule[-2mm]{.5mm}{6mm}_{u_r\ge t}$, for different $t$. Figure 2(d)
presents the experimental moments. It shows, first, quite substantial tails: even when $t=30$ (!),
or greater (in units of $\sigma_u$), the contribution to the cumulative moments is substantial.
Second, we see a remarkable feature: the negative contributions exceed the positive not only
at small $t$, corresponding to the core of the distribution, but also far in tails. 
For comparison, we constructed analogous moments based on $I_r$  (without tails). We see that
these moments are  decaying fast, already for $t>3$, or so, as we would indeed  expect 
from 
(pseudo)-Gaussian distributions.

In \cite{94}, \cite{96} a model was suggested explaining how the asymmetry appears. Figure 2(a)
shows a ramp-structure. Obviously, $\langle\partial_x u(x)\rangle=0$, while
$\langle\partial_x u(x)^3\rangle<0$. In addition, the negative part of $\partial_x 
u(x)$ is certainly intermittent, see Fig. 2(b), and that is how the idea of intermittency 
being connected to the asymmetry came into life.

\begin{figure}[h]
\psfig{file=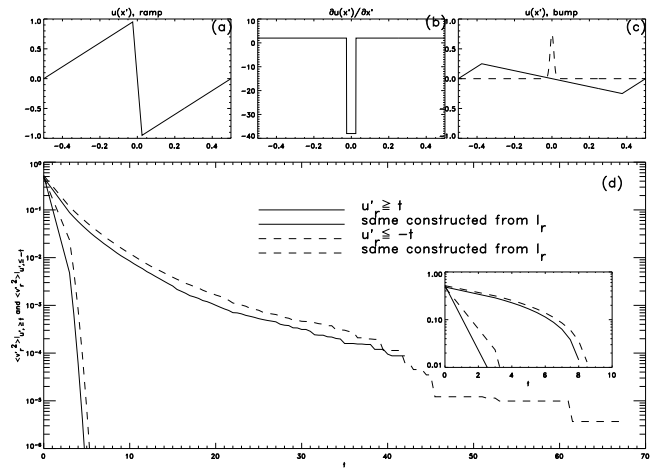,width=3.5in}
\caption{(a) Ramp-structure. (b) Derivative of the ramp.
(c) Bump-structure. (d) Cumulative moments for different
cut-off numbers $t$. The inset corresponds to same cumulative moments, calculated
for the model (\ref{second}).
}
\end{figure}

This model is only heuristic, however. It was shown \cite{1}-\cite{2}  that Burgers vortex, 
embedded into a converging motion, acquires negative skewness, this picture containing both
asymmetry and intermittency. The ramp-model does not exactly
correspond to it. 
More realistic modification of this model is a bump-model \cite{01}, see Fig. 2(c), where $u(x)$ is a
sum of the solid and dashed lines. Here again, $\langle\partial_x u(x)\rangle=0$ while
$\langle\partial_x u(x)^3\rangle<0$. This model simulates a converging motion (solid line in the
vicinity of the dashed peak), naturally generating a vortex (dashed line). Supposedly, this
structure in the longitudinal velocity appears in the vicinity of a Burgers vortex. Both
ramp-model and bump-model are 1D, and therefore they, of course,
 do not reflect any structures appearing in the transverse field.
Therefore, we need further modification of the model to make it 2D, or 3D type. Besides, a real
vortex in  a converging motion depicted by a solid line in Fig. 2(c) would be described by a shear
of $v_y$-component of velocity (rather than generating a shear in $v_x$). If that is the 
case, then  the
 $u_r$-$v_r^2$ anti-correlation will appear. Indeed,
when $u_r<0$ (converging motion), the vortex is generated increasing $v_y^2$, that is, $v_r^2$,
 while for $u_r>0$ 
(diverging motion), the vortex is not generated (and $v_r^2$ is smaller than averaged).

Consider, therefore, a 2D-model: $v_x=f_1(x')$,  corresponding to the solid line in Fig. 2(c), and
$v_y=f_2(x')$,  depicted by  a dashed line. Let
$\alpha=f_1(x'+r_x)-f_1(x')$, and $\omega=f_2(x'+r_x)-f_2(x')$. Then,
\begin{equation}
u_{r}=[\alpha \cos{\phi}+\omega\sin{\phi}]\cos{\phi},~~~
v_{r}=[-\alpha \sin{\phi}+\omega\cos{\phi}]\cos{\phi},
\label{second}
\end{equation}
where $r_x=r\cos{\phi},~~r_y=r\sin{\phi}$. We thus have two averages: over  $x'$, and
over $\phi$. As a result, $\langle u_r\rangle=\langle v_r\rangle=0$, while $\langle u_r^3 \rangle=
\langle \alpha^3\rangle\phi_{60}+3 \alpha(0)\langle\omega^2\rangle\phi_{40}$,
$\langle u_rv_r^2\rangle=\langle \alpha^3\rangle\langle\phi_{40}+\alpha(0)\langle\omega^2\rangle
(\phi_{60}-2
\phi_{40})$, where $\phi_{60}=\langle \cos^6\phi\rangle=5/16,
\phi_{42}=\langle\cos^4\phi\sin^2\phi\rangle=1/16$
(and $\langle v_r^3\rangle=0$). Here we considered small $r$'s, so that $\omega$
is strongly peaked at $x'=0$, and therefore, when combined with $\omega$, the value of $\alpha$
contributes only at $x'=0$. As $\alpha(0)<0$ (converging motion), and $|\omega| \gg |\alpha|$,
both $\langle u_r^3 \rangle$ and $\langle u_rv_r^2\rangle$ are negative.

So far, the model is consistent with the experimental data. The model contains several
free parameters. It is interesting to note that by  choosing them just in a ``reasonable"
way, we
immediately reproduce the real experimental values for $\langle u_r^3\rangle$, and
$\langle u_r v_r^2\rangle$ with a decent accuracy. To do it even better, we used
computer routines to optimize these parameters so that they fit the experimental
values in the best way. 
We now are ready to calculate the
cumulative moments $\langle u_r v_r^2\rangle\rule[-2mm]{.5mm}{6mm}_{u_r\le -t}$, and 
$\langle u_r v_r^2\rangle\rule[-2mm]{.5mm}{6mm}_{u_r\ge t}$, for different $t$, corresponding
to this model. They are shown in the inset of Fig. 2(d). Qualitatively, we see the same features
as in experimental moments. Namely, there are substantial tails, obviously related to the 
presence of the vortex, and the negative part always exceeds the positive one. 

In conclusion we note that the cumulative moments are useful in studying the tails of
the distributions: we thus consider the contribution of the tails, as if the core of the
distribution does not contribute at all. We saw that the Kolmogorov law, and related
third order $u_r$-$v_r^2$ correlation can be satisfactory reproduced by the tails only.
In contrast, the third order moments corresponding to some pseudo-Gaussian distributions are
 poorly reproduced by the contribution
of the tails. As these third order moments do not vanish because of the asymmetry of
the distributions, we assume that the intermittency (i.e.,  substantial contribution
of the tails of the distributions) is related to the asymmetry. This conjecture can be
checked directly, comparing positive and negative  contributions. We see that predicted
difference between the positive and negative parts is present not only at the
core of the distribution, but also stretches   far into the tails.

 The intermittency
related to the asymmetry comes out naturally from the ramp-model, and its 2D modification -- 
the
bump-model. It simply presents a vortex embedded into a converging motion. Some analytical
representation of this model shows quantitatively  the same behavior as the experimental
data. We conclude that it is consistent with the above interpretation of intermittency
related to the asymmetry. As the third moment is a relatively low moment, this
conjecture suggests an useful tool in studying the intermittency of turbulence.

I thank K.R. Sreenivasan, S. Kurien, and R. Rosner for sending me these valuable data, and
for discussions.

\end{document}